# CONTINUOUS ITERATION OF DYNAMICAL MAPS[#]

R. Aldrovandi and L.P. Freitas

INSTITUTO DE FÍSICA TEÓRICA

UNIVERSIDADE ESTADUAL PAULISTA

Rua Pamplona, 145

01405-900  São Paulo  SP Brazil

E-Mails: RA@AXP.IFT.UNESP.BR and LFREITAS@IFT.UNESP.BR.

---

[#] With support of CNPq, Brasilia.

## ABSTRACT

A precise meaning is given to the notion of continuous iteration of a mapping. Usual discrete iterations are extended into a dynamical flow which is a homotopy of them all. The continuous iterate reveals that a dynamic map is formed by independent component modes evolving without interference with each other.

## 1. Introduction

There are two main approaches to describe the evolution of a dynamical system.[1] The first has its roots in classical mechanics – the solutions of the dynamical differential equations provide the continuous motion of the representative point in phase space.[2] The second takes a quite different point of view: it models evolution by the successive iterations of a well-chosen map, so that the state is known after each step, as if the "time" parameter of the system were only defined at discrete values.[3] It is possible to go from the first kind of description to the second through the snapshots leading to a Poincaré map. Our aim here is to present the first steps into a converse procedure, going from a discrete to a continuous description *while preserving the idea of iteration*. This is possible if we are able to "interpolate" between the discrete values in such a way that the notion of iteration keeps its meaning in the intervals.

Iteration is a particular case of function composition: given a basic map $f(x) = f^{<1>}(x)$ , its first iterate is $f^{<2>}(x) = [f \circ f](x) = f[f(x)]$, its n-th iterate is $f^{<n>}(x) = f[f^{<n-1>}(x)] = f^{<n-1>}[f(x)]$, etc. The question is whether or not, given the set of functions $f^{<n>}(x)$, an interpolation $f^{<t>}(x)$ with real values of t can be found which represents the one-parameter continuous group (or semigroup) describing the dynamical flow of the system. In order to do it, $f^{<t>}$ should satisfy the conditions

$$f^{<t>}[f^{<t'>}(x)] = f^{<t'>}[f^{<t>}(x)] = f^{<t+t'>}(x); \qquad (1.1)$$

$$f^{<0>}(x) = Id(x) = x. \qquad (1.2)$$

We shall find in what follows a map interpolation $f^{<t>}(x)$ with these properties. It is a well-known fact that Taylor series $f(x)$, $g(x)$ satisfying the conditions $f(0) = 0$, $g(0) = 0$ and with nonvanishing coefficients of x are invertible and constitute a group by composition.[4] The neutral element is the identity function $e(x) = Id(x) = x$  and two functions f, g are inverse to each other if it is true that $f[g(x)] = g[f(x)] = Id(x) = x$. In this case, $g = f^{<-1>}$ and $f = g^{<-1>}$ (we are taking the liberty of using the word "function" even for purely formal series and multi-valued maps). The clue to the question lies in the formalism of Bell polynomials, which attributes to every such a function f a matrix B[f], whose inverse represents the inverse function and such that the matrix product represents the composition operation. In other words, these matrices provide a linear representation of the group formed by the functions with the operation of composition. Composition is thus represented by matrix product and, consequently, iterations are represented by matrix powers. Furthermore, the representation is faithful, and the function f is completely determined by B[f]. Now, in the matrix group there does exist a clear interpolation of discrete powers by real powers and the inverse way, going from matrices to functions, yields a map interpolation with the desired properties.

Section 2 is a short presentation of Bell polynomials, with only the minimum information necessary to our present objective. It is shown how a matrix B[f] can be found which represents each formal series f, and that the composition $f \circ g$ of two functions is represented by the (right-)product of the respective matrices: $B[f \circ g] = B[g]B[f]$. The identity matrix corresponds to the identity function, $B[Id] = I$, the matrix $B[f^{<n>}]$ corresponding to the n-th iterate $f^{<n>}$ is the n-th power $B^n[f]$ and the Lagrange inverse $f^{<-1>}$ to a series f is represented by the respective inverse matrix, $B[f^{<-1>}] = B^{-1}[f]$. The necessity of finding $B^t[f]$ for non-integer "t" leads to the problem of defining functions of matrices, succinctly discussed in section 3. Given a matrix B, there exists a very convenient basis of projectors in terms of which any function of B is defined in a simple way. A method is given to obtain the members of this basis in a closed form, in terms of powers of B.

The procedure is applied to Bell matrices in section 4 to obtain $B^t$ [f] = $B[f^{<t>}]$ for any value of t, from which the function $f^{<t>}(x)$ can be extracted and shown to satisfy conditions (1.1-2). It turns out that, though it is quite natural to call "time" the continuous label t, this "time" is related to a certain class of flows, amongst all those leading to a specific Poincaré map. There is an extra bonus: the matrix decomposition in terms of projectors is reflected in a decomposition of the original map, and of its iterate, into a sum in terms of certain "elementary functions", each one with an independent and well-defined time evolution.

## 2. Bell matrices

Given a formal series with vanishing constant term,

$$g(x) = \sum_{j=1}^{\infty} \frac{g_j}{j!} \, x^j , \qquad (2.1)$$

its Bell polynomials $B_{nk}[g]$ are certain polynomials[5] in the Taylor coefficients $g_i$, defined by

$$B_{nk}(g_1, g_2, \ldots, g_{n-k+1}) = \frac{1}{k!} \left\{ \frac{d^n}{dt^n} \, [g(t)]^k \right\}_{t=0} . \qquad (2.2)$$

Their properties are in general obtained from their appearance in the multinomial theorem, which reads

$$\frac{1}{k!}\left(\sum_{j=1}^{\infty}\frac{g_j}{j!}t^j\right)^k = \sum_{n=k}^{\infty}\frac{t^n}{n!}\,B_{nk}\,(g_1,\,g_2,\,\ldots,\,g_{n-k+1}). \quad (2.3)$$

Depending on the situation, one or another of the notations

$$B_{nk}[g] = B_{nk}(g_1,\,g_2,\,\ldots,\,g_{n-k+1}) = B_{nk}\{g_j\}\,, \qquad (2.4)$$

is more convenient. The symbol $\{g_j\}$ represents the Taylor coefficient list of $g$, with $g_j$ a typical member. Some properties coming immediately from the multinomial theorem are the following:

$$B_{n1}\,[g] = g_n\,; \qquad\qquad (2.5)$$
$$B_{nn}\,[g] = (g_1)^n\,; \qquad\qquad (2.6)$$
$$B_{nk}\,[cg(t)] = c^k\,B_{nk}\,[g(t)]\,; \qquad\qquad (2.7)$$
$$B_{nk}\,[g(ct)] = c^n\,B_{nk}\,[g(t)]\,, \qquad\qquad (2.8)$$

where $c$ is a constant. Given two formal Taylor series

$$f(u) = \sum_{j=1}^{\infty}\frac{f_j}{j!}\,u^j\,,\; g(t) = \sum_{j=1}^{\infty}\frac{g_j}{j!}\,t^j\,, \qquad (2.9)$$

their composition

$$F(t) = [f \circ g](t) = f[g(t)] = \sum_{j=1}^{\infty}\frac{t^n}{n!}\,F_n[f;\,g] \qquad (2.10)$$

will have the Taylor coefficients $F_n$ given by the Faà di Bruno formula,

$$F_n[f;\,g] = \sum_{k=1}^{n}f_k\,B_{nk}\,\{g_j\}\,. \qquad (2.11)$$

Other properties can be obtained from the double generating function

$$e^{ug(t)} - 1 = \sum_{n=1}^{\infty}\frac{t^n}{n!}\sum_{j=1}^{n}u^j\,B_{nj}[g]\,. \qquad (2.12)$$

Series like (2.1) constitute a group under the composition operation $(f \circ g)(x) = f[g(x)]$. The identity series "e" such that $e(x) = Id(x) = x$ plays the role of the neutral element and each series g possesses an inverse $g^{<-1>}$, satisfying $g^{<-1>} \circ g = g \circ g^{<-1>} = e$ and given by the Lagrange inversion formula. The simplest example of a Bell matrix and its inverse is given by the well-known case of the Stirling numbers: matrices formed by the first and second kind Stirling numbers are inverse to each other, because they correspond to functions inverse to each other. In effect, consider the series

$$g(x) = \ln(1+x) = \sum_{j=1}^{\infty} \frac{(-)^{j-1}}{j} x^j , \qquad (2.13)$$

whose inverse is

$$f(u) = e^u - 1 = \sum_{j=1}^{\infty} \frac{1}{j!} u^j . \qquad (2.14)$$

A generating function for the Stirling numbers of the first kind $s_k^{(j)}$ is

$$\frac{1}{k!} \left( \ln(1+x) \right)^k = \sum_{n=k}^{\infty} \frac{x^n}{n!} s_n^{(k)} . \qquad (2.15)$$

It follows from (2.3) that

$$B_{nk}[\ln(1+x)] = B_{nk}(0!, -1!, 2!, -3!, \dots) =$$

$$= B_{nk}\{(-)^{j-1}(j-1)!\} = s_n^{(k)}. \qquad (2.16)$$

For the Stirling numbers of the second kind $S_k^{(j)}$, the generating function is

$$\frac{1}{k!} (e^u - 1)^k = \sum_{n=k}^{\infty} \frac{u^n}{n!} S_n^{(k)} , \qquad (2.17)$$

from which

$$B_{nk}[e^u - 1] = B_{nk}(1, 1, 1, \dots, 1) = B_{nk}\{1\} = S_n^{(k)}. \qquad (2.18)$$

The inverse property of Stirling numbers is $\sum_{k=j}^{n} s_n^{(k)} S_k^{(j)} = \delta_n^{j}$, the same as

$$\sum_{k=j}^{n} B_{nk}[\ln(1+x)] \; B_{kj}[e^u - 1] = \delta_{nj} . \qquad (2.19)$$

The polynomials $B_{nk}[g]$ are the entries of a (lower-)triangular matrix $B[g]$, with n as row index and k as the column index. From (2.5), the function coefficients constitute the first column, so that actually $B[g]$ is an overcomplete representative of g. From (2.6), the eigenvalues of $B[g]$ are $(g_1)^j$. Triangular matrices form a group, of which the set of matrices $(B_{nk})$ constitutes a subgroup. Hereby comes the most fascinating property of Bell polynomials: the matrices $B[g] = (B_{nk}[g])$, with the operation of matrix product, provide a representation of the series composition group:

$$B\,[g]\,B\,[f] = B\,[f \circ g]\,. \qquad (2.20)$$

It is in reality an anti-representation because of the inverse order, but this does not represent any problem. This property comes easily by using twice (2.3), as

$$\frac{1}{k!}\big(f[g(t)]\big)^k = \frac{1}{k!}\big(\sum_{j=1}^{\infty} \frac{f_j}{j!}g^j\big)^k = \sum_{n=k}^{\infty} \frac{g(t)^n}{n!}\,B_{nk}\,[f] =$$

$$= \sum_{n=k}^{\infty} \sum_{j=n}^{\infty} \frac{t^j}{j!}\,B_{jn}[g]\,B_{nk}\,[f] = \sum_{j=k}^{\infty} \frac{t^j}{j!} \sum_{n=k}^{j} B_{jn}[g]\,B_{nk}\,[f]\,,$$

from which

$$B_{jk}[f(g(t))] = \sum_{n=k}^{j} B_{jn}[g]B_{nk}[f], \qquad (2.21)$$

which is just (2.20). Associativity can be easily checked, and it is trivial to see that the "identity" series $e(x) = x$ has the representative $B_{nk}[e] = \delta_{nk}$, so that $B[e] = I$. Series $g(t)$ with $g_1 = 0$ can be attributed a matrix, but a singular one and, consequently, outside the group. Summing up, infinite Bell matrices constitute a linear representation of the group of invertible formal series. If we consider only the first N rows and columns, what we have is an approximation, but it is important to notice that the group properties hold at each order N. The general aspect of a Bell matrix can be illustrated by the case N = 5:

$$B[g] = \begin{pmatrix} g_1 & 0 & 0 & 0 & 0 \\ g_2 & g_1{}^2 & 0 & 0 & 0 \\ g_3 & 3g_1g_2 & g_1{}^3 & 0 & 0 \\ g_4 & 4g_1g_3+3g_2{}^2 & 6g_1{}^2g_2 & g_1{}^4 & 0 \\ g_5 & 10g_2g_3+5g_1g_4 & 15g_1g_2{}^2+10g_1{}^2g_3 & 10g_1{}^3g_2 & g_1{}^5 \end{pmatrix} . \quad (2.22)$$

The result (2.19) is the best example of the general property

$$B[f]\ B[f^{<-1>}] = I \ . \qquad (2.23)$$

It is evident that, given the series f, its inverse series can be obtained from

$$B[f^{<-1>}] = B^{-1}[f] \qquad (2.24)$$

by simple matrix inversion. The inversion properties of Bell matrices have been used in the study of cluster expansions for real gases.[6] Because $B[g^{<n>}] = B^n[g]$, Bell matrices convert function iteration into matrix power and provide a linearization of the process of iteration.

Suppose now that we are able to obtain the matrix $B^t$, with t an arbitrary real number. The continuous iteration of $g(x)$ will then be that function $g^{<t>}$ such that $B[g^{<t>}] = B^t[g]$. By (2.5), its Taylor coefficients are fixed by $g^{<t>}{}_n = B[g^{<t>}]_{n1} = B^t[g]_{n1}$. To arrive at $B^t$, let us make a short preliminary incursion into the subject of matrix functions.

## 3. Matrix functions

Suppose a function $F(\lambda)$ is given which can be expanded as a power series $F(\lambda) = \Sigma_{k=0}^{\infty} c_k(\lambda - \lambda_o)^k$ inside the convergence circle $|\lambda - \lambda_o| < r$. Then the function $F(B)$, whose argument is now a given $N \times N$ matrix $B$, is defined by $F(B) = \Sigma_{k=0}^{\infty} c_k(B - \lambda_o)^k$ and has a sense whenever the eigenvalues of $B$ lie within the convergence circle. Given the eigenvalues $x_1, x_2, \ldots , x_N$, the set of eigenprojectors $\{Z_j[B] = |x_j\rangle\langle x_j|\}$ constitutes a basis in which $B$ is written

$$B = \sum_{j=1}^{N} x_j \, Z_j[B] \, , \qquad (3.1)$$

and the function $F(B)$, defined as above, can also be written[7] as the matrix

$$F(B) = \sum_{j=1}^{N} F(x_j) \, Z_j[B] \, . \qquad (3.2)$$

Thus, for example, $e^B = \Sigma_{j=1}^{N} e^{x_j} Z_j$ and $B^{\alpha} = \Sigma_{j=1}^{N} x_j{}^{\alpha} Z_j[B]$. The basis $\{Z_j[B]\}$ depends on $B$, but is the same for every function $F$. The $Z_j$'s, besides being projectors (that is, idempotents, $Z^2{}_j = Z_j$), can be normalized so that tr $(Z_j) = 1$ for each $j$ and are then orthonormal by the trace, $\mathrm{tr}(Z_i Z_j) = \delta_{ij}$. Other properties follow easily, for example $\mathrm{tr}[F(B)] = \Sigma_{j=1}^{N} F(x_j) Z_j$ and $\mathrm{tr}[B^k Z_j] = (x_j)^k$. If $B$ is a normal matrix diagonalized by a matrix $U$, $UBU^{-1} = B_{\mathrm{diagonal}}$, then the entries of $Z_k$ are given by $(Z_k)_{rs} = U^{-1}{}_{rk} U_{ks}$ (no summation, of course).

A set of N powers of B is enough to fix the projector basis. Using for F(B) in (3.2) the power functions $B^0 = I$, $B^1$, $B^2$, . . . , $B^{N-1}$, we have $I = \Sigma_{j=1}^{N} Z_j$; $B = \Sigma_{j=1}^{N} x_j Z_j$; $B^2 = \Sigma_{j=1}^{N} x_j^2 Z_j$; . . . ; $B^k = \Sigma_{j=1}^{N} x_j^k Z_j$; . . . ; $B^{N-1} = \Sigma_{j=1}^{N} x_j^{N-1} Z_j$. For $k \geq N$, the $B^k$'s are no more independent. This comes from the Cayley-Hamilton theorem,[8] by which B satisfies its own secular equation

$$\Delta(x) = \det [x\, I - B] = (x-x_1)(x-x_2)(x-x_3) \ldots (x-x_N) = 0.$$

$\Delta(B) = 0$ will give $B^N$ in terms of lower powers of B, so that the higher powers of B can be computed from the lower powers.

Inversion of the above expressions for the powers of B in terms of the $Z_j$'s leads to a closed form for each $Z_j$,

$$Z_j[B] = \frac{(B - x_1)(B - x_2) \ldots (B - x_{j-1})(B - x_{j+1}) \ldots (B - x_{N-1})(B - x_N)}{(x_j - x_1)(x_j - x_2) \ldots (x_j - x_{j-1})(x_j - x_{j+1}) \ldots (x_j - x_{N-1})(x_j - x_N)} \; . \quad (3.3)$$

The function F(B) is consequently given by

$$F(B) = \Sigma_j \left\{ \prod_{k \neq j} \frac{B - x_k}{x_j - x_k} \right\} F(x_j) \, . \quad (3.4)$$

Thus, in order to obtain F(B), it is necessary to find the eigenvalues of B and the detailed form of its first (N-1) powers. Though for N not too large the $Z_j[B]$'s can be directly computed, we shall give closed expressions for them. These expressions involve some symmetric functions of the eingenvalues.

Let us examine the spectrum $\{x_k\}$ of B in some more detail. The eigenvalues $x_k$ will be called "letters" and indicated collectively by the "alphabet" $\mathbf{x} = \{x_1, x_2, x_3, \ldots, x_N\}$. A monomial is a "word". It will be convenient to consider both the alphabet $\mathbf{x}$ and its "reciprocal", the alphabet $\mathbf{x}^* = \{x^*_1, x^*_2, x^*_3, \ldots, x^*_N\}$ where each $x^*_j = -1/x_j$. Notice that taking the reciprocal is an involution, $\mathbf{x}^{**} = \mathbf{x}$. A symmetric function in the variables $x_1, x_2, x_3, \ldots, x_N$ is any polynomial $P(x_1, x_2, x_3, \ldots, x_N)$ which is invariant under all the permutations of the $x_k$'s. Only one kind of them will be needed here, the "j-th elementary symmetric functions", $\sigma_j$ = sum of all words with j distinct letters:

$\sigma_0[\mathbf{x}] = 1$ (by convention)
$\sigma_1[\mathbf{x}] = x_1 + x_2 + x_3 + \ldots + x_N$ ;
$\sigma_2[\mathbf{x}] = x_1 x_2 + x_1 x_3 + \ldots + x_1 x_N + \ldots + x_2 x_3 + x_2 x_4 + \ldots + \ldots + x_{N-1} x_N$ ;
  . . .
$\sigma_N[\mathbf{x}] = x_1 x_2 x_3 \ldots x_{N-1} x_N$ .

The symmetric functions of $\mathbf{x}$ and $\mathbf{x}^*$ are related by

$$(-)^j \sigma_{N-j}[\mathbf{x}] = \sigma_N[\mathbf{x}] \sigma_j[\mathbf{x}^*]. \tag{3.5}$$

Their generating function is

$$\sum_{j=0}^{N} \sigma_j[\mathbf{x}] t^j = \prod_{j=1}^{N} (1 + x_j t) = \prod_{j=1}^{N} (1 - t/x^*_j) = \frac{1}{\sigma_N[\mathbf{x}^*]} \prod_{j=1}^{N} (x^*_j - t) \quad ,$$

so that $\prod_{j=1}^{N} (x^*_j - t) = \sigma_N[\mathbf{x}^*] \sum_{j=0}^{N} \sigma_j[\mathbf{x}] t^j$ . We use the involution property and

(3.5) to write the general expression

$$\prod_{j=1}^{N} (x_j - t) = \sigma_N[\mathbf{x}] \sum_{j=0}^{N} \sigma_j[\mathbf{x}^*] t^j = \sum_{j=0}^{N} (-)^j \sigma_{N-j}[\mathbf{x}] t^j \quad . \tag{3.6}$$

The j-th eigenvalue is absent in the numerator of expression (3.3) for $Z_j$. We shall need some results involving an alphabet with one missing letter. Let $\sigma_{ji}[\mathbf{x}]$ be the sum of all j-products of the alphabet $\mathbf{x}$, but excluding $x_i$. For example, $\sigma_{Ni}[\mathbf{x}] = \Pi_{k \neq i}^{N} x_k$. We put by convention $\sigma_{0i} = 1$ and find that

$$\sigma_{ki}[\mathbf{x}] = \sum_{p=0}^{k} {}_{(-)}{}^p x_i{}^p \sigma_{k-p}[\mathbf{x}] = \sum_{j=0}^{k} (-x_i)^{k-j} \sigma_j[\mathbf{x}] \ . \qquad (3.7)$$

In the absence of the i-th letter, (3.6) becomes

$$\prod_{j=1; j \neq i}^{N} (x_j - t) \ = \sigma_{Ni}[\mathbf{x}] \sum_{j=0}^{N} \sigma_{ji}[\mathbf{x}*]t^j \ . \qquad (3.8)$$

The projectors (3.3) are then

$$Z_j[B] = \prod_{k=1; k \neq j}^{N} \frac{x_k - B}{x_k - x_j} = \frac{\displaystyle\sum_{k=0}^{N} \sigma_{kj}[\mathbf{x}*] \, B^k}{\displaystyle\sum_{k=0}^{N} \sigma_{kj}[\mathbf{x}*]x_j{}^k} \quad , \qquad (3.9)$$

clearly written in the basis $\{I, B, B^2, \ldots, B^{N-1}\}$. In our application to Bell matrices, it will be convenient to use instead the basis $\{B, B^2, \ldots, B^N\}$. This is due to the fact that we shall prefer to start from the matrix B[g], corresponding to g, and not from the matrix I, corresponding to the identity function. The mappings of interest, like the logistic map for example, have a general "$\cap$" aspect and cannot be obtained continuously from the identity map: the identity map has Brower[9] degree 1, while the dynamical maps have degree 0. The Cayley-Hamilton theorem implies $\Sigma_{j=0}^{N}$ $\sigma_j[\mathbf{x}*] \, B^j = 0$, from which we obtain the identity $B^0$ as

$$I = - \sum_{j=1}^{N} \sigma_j[\mathbf{x}^*] \ B^j \quad . \tag{3.10}$$

Replacing this identity in (3.9), the projectors are recast into another form,

$$Z_i[B] = \frac{I + \sum_{k=1}^{N} \sigma_{ki}[\mathbf{x}^*] \ B^k}{1 + \sum_{k=1}^{N} \sigma_{ki}[\mathbf{x}^*] x_i^k} = \frac{\sum_{k=1}^{N} \{\sigma_{ki}[\mathbf{x}^*] - \sigma_k[\mathbf{x}^*]\} \ B^k}{\sum_{k=1}^{N} \{\sigma_{ki}[\mathbf{x}^*] - \sigma_k[\mathbf{x}^*]\} x_i^k} \quad . \tag{3.11}$$

Using (3.7) and (3.5) we get

$$Z_i[B] = \frac{\sum_{k=1}^{N} \{\sum_{j=0}^{k-1} (x_i)^{j-k} \sigma_j[\mathbf{x}^*]\} B^k}{\sum_{j=0}^{N-1} (N-j)(x_i)^j \sigma_j[\mathbf{x}^*]} = \frac{\sum_{k=1}^{N} \{\sum_{j=0}^{k-1} (x_i)^{j-k} (-)^j \sigma_{N-j}[\mathbf{x}]\} B^k}{\sum_{j=0}^{N-1} (N-j)(x_i)^j (-)^j \sigma_{N-j}[\mathbf{x}]} \quad , \tag{3.12}$$

where we have also used $\Sigma_{r=1}^{N}\{\Sigma_{j=0}^{r-1} (x_i)^j \sigma_j[\mathbf{x}^*]\} = \Sigma_{j=0}^{N-1} (N-j)(x_i)^j \sigma_j[\mathbf{x}^*]$. There is a good immediate check: replacing B by the eigenvalues we find, as expected,

$$Z_i[x_k] = \delta_{ik} I. \tag{3.13}$$

The projectors are now clearly in the basis $\{B, B^2, \ldots, B^N\}$. Actually, each $Z_i$ is now just that given in the basis $\{I, B, B^2, \ldots, B^{N-1}\}$ multiplied by $(B/x_i)$: instead of (3.3),

$$Z_j[B] = \frac{(B - x_1)(B - x_2) \ldots (B - x_{j-1}) \ B \ (B - x_{j+1}) \ldots (B - x_{N-1})(B - x_N)}{(x_j - x_1)(x_j - x_2) \ldots (x_j - x_{j-1}) \ x_j \ (x_j - x_{j+1}) \ldots (x_j - x_{N-1})(x_j - x_N)} \quad . \tag{3.14}$$

The $Z_i$'s and the powers $B^k$ can be seen as components of two formal column "vectors". The linear conditions $B^n = \sum_{j=1}^{N} x_j{}^n Z_j$ are then represented by a matrix $\Lambda = [x_j{}^n]$,

$$
\begin{pmatrix} B^1 \\ B^2 \\ \cdot \\ \cdot \\ \cdot \\ B^n \end{pmatrix} = \begin{pmatrix} x_1 & x_2 & x_3 & \cdot & \cdot & x_N \\ x_1{}^2 & x_2{}^2 & x_3{}^2 & \cdot & \cdot & x_N{}^2 \\ \cdot & \cdot & \cdot & \cdot & \cdot & \cdot \\ \cdot & \cdot & \cdot & \cdot & \cdot & \cdot \\ \cdot & \cdot & \cdot & \cdot & \cdot & \cdot \\ x_1{}^n & x_2{}^n & x_3{}^n & \cdot & \cdot & x_N{}^n \end{pmatrix} \begin{pmatrix} Z_1 \\ Z_2 \\ \cdot \\ \cdot \\ \cdot \\ Z_n \end{pmatrix} , \quad (3.15)
$$

and what we have done has been to obtain its inverse:

$$
Z_i[B] = \sum_{k=1}^{N} [\Lambda^{-1}]_{ik} B^k; \quad (3.16)
$$

$$
[\Lambda^{-1}]_{ik} = \frac{\sum_{j=0}^{k-1} (x_i)^{j-k}(-)^j \sigma_{N-j}[\mathbf{x}]}{\sum_{j=0}^{N-1} (N-j)(x_i)^j(-)^j \sigma_{N-j}[\mathbf{x}]} \quad . \quad (3.17)
$$

It seems a difficult task to improve the above expressions, as it would mean knowing a closed analytical expression for the recurrent summation of the form $\sum_{j=0}^{k-1} u^j \sigma_j[\mathbf{x}^*]$. A closed expression for $\sigma_j[\mathbf{x}^*]$ would be necessary and, even for the simple alphabet consisting of powers of a fixed letter a, which we shall find in the application to Bell matrices, this would be equivalent to solving an as yet unsolved problem in Combinatorics. In effect, in terms of such an alphabet $\{a^j\}$ with N letters, the symmetric function is given by $\sigma_k = \sum_{j=1}^{N(N+1)/2} q_{j,k,N} a^j$, where $q_{j,k,N}$ = number of partitions of j into k unequal summands, each one $\leq N$. These conditional partition numbers have the generating function $\Pi_{r=1}^{N}[1+ua^r] = 1 + \Sigma_{j\&k\geq 1}$ $q_{j,k,N}a^j u^k$, but have no known closed expression. They are calculated, one by one, just in this way. [10]

Bell matrices are not normal, that is, they do not commute with their transposes. Normality is the condition for diagonalizability. This means that Bell matrices cannot be put into diagonal form by a similarity transformation. As it happens, this will not be a difficulty because we know their eigenvalues. That functions of matrices are completely determined by their spectra is justified on much more general grounds. Matrix algebras are very particular kinds of von Neumann algebras and it is a very strong result of the still more general theory of Banach algebras[11] that functions on such spaces, as long as they can be defined, are fixed by the spectra. Another point worth mentioning is that the infinite Bell matrices which constitute the true, complete representation of the group of invertible series will belong, as limits for N $\rightarrow \infty$ of N×N matrices, to a hyperfinite von Neumann algebra. Our considerations here are purely formal from the mathematical point of view, as we are only discussing formal series. We are not concerned with the topological intricacies involved in the convergence problems, though they surely deserve a detailed study. By the way, the infinite algebra generated by Bell matrices would provide a good guide in the study of function algebras with the composition operation.

## 4. The continuous iterate

We are now in condition to find, given a function g, the matrix $B^t$ [g] and its corresponding function, the continuous iterate $g^{<t>}(x)$. As many things – the B[g] spectrum for example – will depend only on the first-order Taylor coefficient, we shall put $g_1 = a$. By (2.6), the letters in the eigenvalue-alphabet of $B[g]$ will be simple powers of a and the alphabet $\mathbf{a} = (a^1, a^2, ...., a^N)$ will have the reciprocal $\mathbf{a}^* = (-a^{-1}, -a^{-2}, ...., -a^{-N})$. The matrix $\Lambda$ has entries $(\Lambda_{ik}) = (a^{ik})$ and the projectors

$$Z_i[B] = \sum_{k=1}^{N} \Lambda^{-1}{}_{ik} B^k[g] \tag{4.1}$$

have now the coefficients

$$[\Lambda^{-1}]_{ik} = \frac{\displaystyle\sum_{j=0}^{k-1} a^{i(j-k)}(-)^j \sigma_{N-j}[\mathbf{a}]}{\displaystyle\sum_{r=1}^{N}\sum_{j=0}^{r-1} a^{ij}(-)^j \sigma_{N-j}[\mathbf{a}]} \quad .$$

We can verify easily that $\mathrm{tr}\, Z_i[B] = 1$ and $BZ_i = a^i Z_i$. A consequence of the latter is

$$f(B)Z_i = f(a^i)Z_i \ , \tag{4.2}$$

whose particular case

$$B^t Z_i = a^{it} Z_i \tag{4.3}$$

will be helpful later on. To give an idea of their aspect, we show the projector matrices $Z_i^{(N)}[B[g]]$ for $N = 3$:

$$Z_1^{(3)} = \begin{pmatrix} 1 & 0 & 0 \\ \dfrac{g_2}{g_1(1-g_1)} & 0 & 0 \\ \dfrac{3g_2{}^2 + g_3(1-g_1)}{g_1(1-g_1)^2(1+g_1)} & 0 & 0 \end{pmatrix} ; \quad Z_2^{(3)} = \begin{pmatrix} 0 & 0 & 0 \\ \dfrac{-g_2}{g_1(1-g_1)} & 1 & 0 \\ \dfrac{-3g_2{}^2}{g_1{}^2(1-g_1)^2} & \dfrac{3g_2}{g_1(1-g_1)} & 0 \end{pmatrix} ;$$

$$Z_3^{(3)} = \begin{pmatrix} 0 & 0 & 0 \\ 0 & 0 & 0 \\ \dfrac{3g_2{}^2 - g_1 g_3 (1-g_1)}{g_1{}^2 (1-g_1)^2 (1+g_1)} & \dfrac{-3g_2}{g_1(1-g_1)} & 1 \end{pmatrix} .$$

As they have not the form (2.22), they cannot be the Bell matrices of any function. They inherit, however, a good property of the Bell matrices: for each N, the projectors $Z_j^{(N)}$ contain, in their higher rows, the projectors $Z_j^{(k)}$ for all $k < N$. The upper-left 2×2 submatrices of the $Z_k^{(3)}$'s above are just $Z_k^{(2)}$.

We can take the first-column entries of the matrix (4.1) as Taylor coefficients defining the functions

$$R_i^{(N)}(x) = \sum_{r=1}^{\infty} \frac{x^r}{r!} \, [Z_i[B]]_{r1} \qquad (4.4)$$

$$= \sum_{k=1}^{N} \left[ \Lambda^{-1}{}_{ik} g^{<k>}(x) \right] . \qquad (4.5)$$

To each projector $Z_i$ corresponds such an "elementary function" $R_i^{(N)}(x)$, a relationship reflecting in part that between the series and their Bell matrices. Taking the summation of all $R_i^{(N)}$'s in (4.4) and using $I = \Sigma_{i=1}^{N} Z_i$, we find

$$\sum_{i=1}^{N} R_i^{(N)}(x) = \sum_{r=1}^{\infty} \frac{x^r}{r!} \, [\sum_{i=1}^{N} Z_i[B]]_{r1} = \sum_{r=1}^{\infty} \frac{x^r}{r!} \, \delta_{r1} = x. \quad (4.6)$$

Thus, just as the projectors give a decomposition of the identity matrix, the elementary functions provide a decomposition of the identity function,

$$Id = \sum_{i=1}^{N} R_i{}^{(N)} \, .$$
(4.7)

Applying to $B[g]$ the general form (3.2) for the function of a matrix, we have

$$B^t = \sum_{i=1}^{N} a^{it} Z_i[B] = \sum_{k=1}^{N} C^{(N)}{}_k(t) \, B^k \, ,$$
(4.8)

where

$$C^{(N)}{}_k(t) = \sum_{i=1}^{N} a^{it} \Lambda^{-1}{}_{ik} \, .$$
(4.9)

The coefficients of the continuum iterate function

$$g^{<t>}(x) = \sum_{r=1}^{\infty} g_r{}^{<t>} \frac{x^r}{r!}$$
(4.10)

will be

$$g_r{}^{<t>} = [B^t[g]]_{r1} = \sum_{i=1}^{N} a^{it} [Z_i[B]]_{r1} = \sum_{k=1}^{N} C^{(N)}{}_k(t) \, B^k[g] \,]_{r1}$$

$$= \sum_{k=1}^{N} C^{(N)}{}_k(t) \, g_r{}^{<k>} .$$
(4.11)

Therefore,

$$g^{<t>}(x) = \sum_{k=1}^{N} C^{(N)}{}_k(t) \, g^{<k>}(x) = \sum_{k=1}^{N} \Big[ \sum_{i=1}^{N} a^{it} \Lambda^{-1}{}_{ik} \Big] g^{<k>}(x) \, .$$
(4.12)

Time dependence is factorized in the alternative form

$$g^{<t>}(x) = \sum_{k=1}^{N} a^{kt} R_k{}^{(N)}(x).$$
(4.13)

One of the announced properties, (1.2), follows immediately: $g^{<0>}(x) = Id(x)$. For $t = 1$, a suggestive decomposition of the function comes up:

$$g(x) = \sum_{k=1}^{N} a^k \, R_k^{(N)}(x). \tag{4.14}$$

It remains to show that (4.13) does satisfy property (1.1). Notice first that, from the definition (4.4) of $R_i^{(N)}$ and the multinomial theorem (2.3),

$$R_i^{(N)}[g^{<t>}(x)] = \sum_{r=1}^{\infty} [Z_i[B]]_{r1} \frac{[g^{<t>}(x)]^r}{r!} = \sum_{r=1}^{\infty} \sum_{j \geq r} \frac{x^j}{j!} B_{jr}[g^{<t>}] [Z_i[B]]_{r1} =$$

$$\sum_{j=1}^{\infty} \frac{x^j}{j!} \sum_{r=1}^{j} B_{jr}^t[g][Z_i[B]]_{r1} = \sum_{j=1}^{\infty} \frac{x^j}{j!} [B^t Z_i]_{j1} . \tag{4.15}$$

Therefore,

$$g^{<t'>}(g^{<t>}(x)) = \sum_{i=1}^{N} a^{it'} R_i^{(N)}[g^{<t>}(x)] = \sum_{r=1}^{\infty} \frac{x^r}{r!} \sum_{i=1}^{N} a^{it'} [B^t Z_i]_{r1} =$$

$$= \sum_{r=1}^{\infty} \frac{x^r}{r!} [B^t \sum_{i=1}^{N} a^{it'} Z_i]_{r1} = \sum_{r=1}^{\infty} \frac{x^r}{r!} [B^t B^{t'}]_{r1} = \sum_{r=1}^{\infty} \frac{x^r}{r!} g^{<t+t'>}_{r1} = g^{<t+t'>}(x),$$

just the result looked for. Notice also that, using (4.3) and (4.4) in (4.15), we obtain

$$R_k^{(N)}[g^{<t>}(x)] = a^{kt} R_k^{(N)}(x) \tag{4.16}$$

and, consequently,

$$R_k^{(N)}[g^{<t+t'>}(x)] = a^{k(t+t')} R_k^{(N)}(x) = a^{kt'} R_k^{(N)}[g^{<t>}(x)] . \tag{4.17}$$

The function decomposition is preserved in time, as we can write

$$g(x) = \sum_{k=1}^{N} R_k^{(N)}[g(x)] \ ; \ g^{<t>}(x) = \sum_{k=1}^{N} R_k^{(N)}[g^{<t>}(x)]. \tag{4.18}$$

All the above expressions hold for each value of $N$ and give approximations to that order. Of course, exact results would only really come out when N → ∞. The exact expressions would, as long as they are defined, take forms like

$$R_j(x) = \sum_{k \geq 1} \left[ \Lambda^{-1}{}_{jk} g^{<k>}(x) \right];$$ (4.19)

$$g(x) = \sum_{k \geq 1} a^k R_k(x);$$ (4.20)

$$g^{<t>}(x) = \sum_{k \geq 1} a^{kt} R_k(x).$$ (4.21)

It is tempting to conjecture that the elementary functions $R_i$ play, in function space and with the operation of composition, the role of projectors. In fact, this is not so. Their properties mirror only partially those of the projectors $Z_i$. If we calculate $R_i [R_j(x)]$ using (4.4) and the multinomial theorem, we find $R_i [R_j(x)] = R_i(x^j/j!)$. The composition gives rise to a change in the variable. Thus, the functions $R_i$ fail even to satisfy the defining idempotent property: in general, $R_i \circ R_i \neq R_i$.

We can introduce $\varepsilon = \ln a$ and rewrite (4.13) as

$$x(t) = g^{<t>}(x_o) = \sum_{k \geq 1} e^{k \varepsilon t} R_k(x_o) ,$$ (4.22)

a decomposition of $g^{<t>}(x)$ into a sum of modes, each one evolving independently according to

$$x_k(t) = e^{k \varepsilon t} x_k(0) = e^{k \varepsilon t} R_k(x_o).$$ (4.23)

The imaginary "frequency" $k\varepsilon$ plays the role of a "modular Lyapunov exponent" of the k-th mode. We have thus a "multi-hamiltonian" flow x(t) with one hamiltonian for each projector component: equations (4.16,17) show $R_k^{(N)}$ as a representation of the one-dimensional group engendered by the k-th dynamical flow

$$R_k[g^{<t>}(x)] = e^{k\epsilon t} R_k[g^{<0>}(x)] \ . \qquad (4.24)$$

The function $g^{<t>}(x)$ is actually a Lagrange interpolation in the convenient variable $(g_1)^t = a^t$, which coincides with the discrete iterates at each integer value of t and keeps the meaning of a continuum iterate in between. It is a homotopy of all the usual discrete iterations. As announced in Section 3, we have used the basis $\{B, B^2, B^3, \ldots, B^N\}$, instead of $\{I, B, B^2, \ldots, B^{N-1}\}$, because in the latter the matrix "I" would correspond to the identity function $Id(x) = x$, which does not belong to the same homotopy class.

The variable t has been given the sense of a "time". If it is really time, $g^{<n>}(x)$ will give the n-th point of a Poincaré map. Now, there are in principle (infinitely) many dynamical flows corresponding to a given Poincaré map. The function $g^{<t>}(x)$ as given above would correspond to a class of them, that of the flows with equal intervals of time between successive points on the Poincaré section. To consider other cases, we recall that any strictly monotonous function of a first-given "time" is another "time".

We cannot resist exhibiting a last "thing of beauty". By the very manner it has been obtained, the expression (4.12) for the continuum iterate is equivalent to the highly mnemonic determinantal equation

$$\begin{vmatrix} g^{<1>}(x) & a & a^2 & \ldots & \ldots & a^N \\ g^{<2>}(x) & a^2 & a^4 & \ldots & \ldots & a^{2N} \\ g^{<3>}(x) & a^3 & a^6 & \ldots & \ldots & a^{3N} \\ \ldots & \ldots & \ldots & \ldots & \ldots & \ldots \\ g^{<N>}(x) & a^N & a^{2N} & \ldots & \ldots & a^{N2} \\ g^{<t>}(x) & a^t & a^{2t} & \ldots & \ldots & a^{Nt} \end{vmatrix} = 0. \qquad (4.25)$$

Expansion along the first column or the last row and comparison with (4.11) and (4.13) will give determinant expressions for $C^{(N)}_k(t)$ and $R_k^{(N)}(x)$.

## 5. Concluding remarks

In the passage from functions to Bell matrices, composition is translated into matrix product and iteration into matrix powers. Continuous powers of matrices have a sound meaning, which is translated back into continuous iteration and makes of it also a sound concept, respecting properties (1.1-2). The usual discrete iterations are thus extended into a continuous flow. The term-by-term factorization of the x- and t-dependencies of $g^{<t>}(x)$ reveals itself as a decomposition into independently evolving modes, one for each projector of the corresponding Bell matrix.

Once it is established that iteration can have a continuous meaning and is, furthermore, a homotopy, new possibilities are made open to study, such as the use of the conservation of degrees like Brower's and Morse's. It is our contention that, with the continuum version of iterations, it will be possible to get a better understanding of the detailed unfolding of bifurcations and of the general relationship between the differential and the mapping approaches to chaotic dynamics.